\documentclass[showpacs,twocolumn,aps,floatfix,superscriptaddress,noshowpacs,prl]{revtex4-1}
\usepackage[mathlines]{lineno}

\usepackage{amsthm,amsmath,amssymb,graphicx,bm,color,mathrsfs,verbatim,epstopdf,dcolumn,dsfont,cancel,subfigure}

\usepackage{algorithm}
\usepackage[noend]{algpseudocode}
\makeatletter
\def\BState{\State\hskip-\ALG@thistlm}
\makeatother

\usepackage{bbold} 
\usepackage{graphicx}

\graphicspath{{./},{../figs_prl/},{./figs_prl/}}

\usepackage{hyperref}
\hypersetup{ 
	colorlinks = true
}

\usepackage{color}

\begin{document}


\title{Prethermalization and Thermalization in Periodically-Driven Many-Body Systems \\
	away from the High-Frequency Limit}%

\author{Christoph Fleckenstein}
\email{christoph.fleckenstein@physik.uni-wuerzburg.de}
\affiliation{Universit\"at W\"urzburg, Am Hubland, 97074 W\"urzburg, Germany}
\affiliation{Department of Physics, KTH Royal Institute of Technology, SE-106 91 Stockholm, Sweden}
\author{Marin Bukov}
\email{mgbukov@phys.uni-sofia.bg}
\affiliation{Department of Physics, University of California, Berkeley, CA 94720, USA}
\affiliation{Department of Physics, St.~Kliment Ohridski University of Sofia, 5 James Bourchier Blvd, 1164 Sofia, Bulgaria}

\begin{abstract}
We investigate a class of periodically driven many-body systems that allows us to extend the phenomenon of prethermalization to the vicinity of isolated intermediate-to-low drive frequencies away from the high-frequency limit. We provide numerical evidence for the formation of a parametrically long-lived prethermal plateau, captured by an effective Floquet Hamiltonian computed using the replica inverse-frequency expansion, and demonstrate its stability w.r.t.~random perturbations in the drive period. 
Considering exclusively nonintegrable Floquet Hamiltonians, we find that heating rates are non-universal: we observe Fermi's Golden Rule scaling, power-law scaling inconsistent with the Golden Rule, and non-power law scaling, depending on the drive.
Despite the asymptotic character of the inverse-frequency expansion, we show that it describes the thermostatic properties of the state all along the evolution up to infinite temperature, with higher-order terms improving the accuracy. 
Our results suggest a dynamical mechanism to  gradually increase temperature in isolated quantum simulators, such as ultracold atoms, and open up an alternative possibility to investigate thermal phase transitions and the interplay between thermal and quantum criticality using Floquet drives. 
\end{abstract} 

\date{\today}
\maketitle

Periodically-driven (Floquet) systems bring a new set of tools to study quantum phenomena~\cite{goldman2014periodically,goldman2015case,eckardt2017atomic,bukov2015universal,rodriguez2020moir,sen2021analytic}. A prominent example is Floquet engineering -- the use of periodic modulation to ascribe new properties to static systems. Experimentally, periodic drives are used to 
engineer topological properties in photonic insulators~\cite{rechtsman_13,hafezi_14,mittal_14}, or
simulate artificial magnetic fields~\cite{struck_13,aidelsburger_13,miyake_13,jotzu_15,nascimbene2015dynamic,  price2017synthetic,tarnowski2019measuring,gorg2019realization}, $\mathbb{Z}_2$-lattice gauge theories~\cite{schweizer2019floquet,barbiero2019coupling}, and strongly-interacting models~\cite{gorg2018enhancement,sandholzer2019quantum} using ultracold atoms and quantum solids~\cite{topp2019topological, mciver2020light, nuske2020floquet}. 
Besides the experimental emulation of unexplored exotic static states, periodic modulations have also brought novel, truly dynamic phenomena, revealing phases of matter without equilibrium counterparts~\cite{else_16,khemani_16,yao_17,quelle2017driving,haldar2018onset,else2020long,wintersperger2020realization,machado2020Long, wintersperger2020realization,mukherjee2020dynamics,mukherjee2020collapse}.

\begin{figure}[t!]
	\includegraphics[width=0.8\columnwidth]{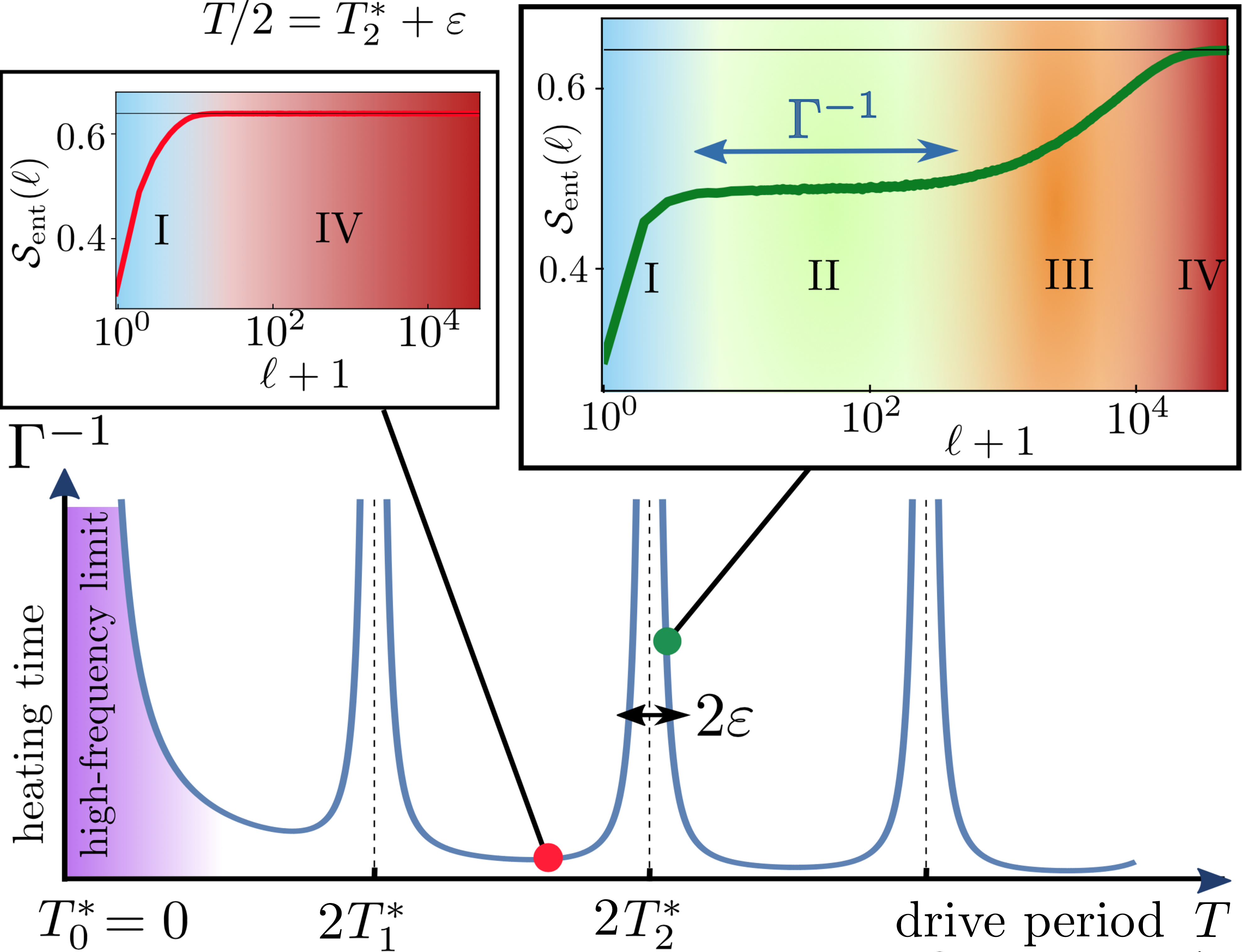}
	\caption{\label{fig:schematic}
		(Schematic) Heating times $\Gamma^{-1}$ against the drive period $T$ for the class of models in Eq.~\eqref{eq:UF_ast}. A prethermal plateau appears in $\varepsilon$-vicinity of commensurate points $T^*_k$, shown in the time evolution of the entanglement entropy $\mathcal{S}_{\mathrm{ent}}(\ell)$ at stroboscopic times $\ell$ [insets]:
		(I) an initial transient of constrained thermalization is followed by (II) a prethermal plateau with subsequent (III) unconstrained thermalization leading to (IV) a featureless infinite temperature state. 
	}
\end{figure}

Conceptually, periodically driven systems define a useful framework to systematically address phenomena in nonequilibrium dynamics, such as equilibration and thermalization~\cite{dalessio2016quantum,moessner2017equilibration}. For drive frequencies much larger than the energy of single-particle processes in the non-driven Hamiltonian, the dynamics of Floquet systems can be divided into four overlapping stages: 
(I) the initial constrained thermalization and the following
(II) prethermal plateau, which lasts exponentially longer with increasing drive frequency, are both captured by a local approximate effective Hamiltonian $H_\mathrm{eff}$~\cite{de2019very}. Eventually, lack of energy conservation sets in, and 
(III) unconstrained thermalization drives the state to 
(IV) infinite-temperature~\cite{dalessio_14,lazarides_14,bar2017absence,weidinger2017floquet,prosen_98a, prosen_99, dalessio2013many,kai2018suppression,abanin_15,mori_15,Avdoshkin2020}.
It was demonstrated that similar prethermal plateaus can be universally protected by topological effects~\cite{lindner2017universal,gulden2020exponentially}, and have been shown to occur also in quasi-periodically driven systems~\cite{dumitrescu2018logarithmically,else2020long,zhao2020random}, nonlocal systems such as the Floquet SYK model~\cite{kuhlenkamp2020periodically}, and for random dipolar driving~\cite{zhao2020random}. A prethermal plateau has also been found to form in isolated (semi-)classical Floquet systems, implying that the underlying mechanism is not governed by quantum effects~\cite{notarnicola2018from,rajak2018stability, howell2019asymptotic, mori2018floquet, rajak2019characterizations, huveneers2020prethermalization,torre2020statistical}. 
At lower drive frequencies, a proliferation of Floquet many-body resonances~\cite{bukov_15_res} causes the system to absorb energy, and prethermalization gives in to immediate heating to a featureless infinite-temperature state.

In this work, we propose a new extension of the notion of prethermalization into the intermediate-to-low frequency regime using a family of step-driven Hamiltonians, for which energy conservation holds exactly for a series of increasing commensurate driving periods $T^\ast_k$. 
In a finite vicinity around $T_k^{\ast}$, as measured by a small number $0\leq\varepsilon<1$, we observe numerically the four stages of thermalization [Fig.~\ref{fig:schematic}] familiar from the high-frequency limit, and show that an effective Hamiltonian computed using the replica inverse-frequency expansion~\cite{vajna2018replica} provides an analytical description. 
We then analyze the $\varepsilon$-dependence of the timescales required to reach the infinite-temperature state: for nonintegrable drives, we show a Fermi Golden Rule scaling implying the durations of the prethermal plateau scales as $\varepsilon^2$. Interestingly, for integrable drives, we find exceptions to the predictions of the Golden Rule, leading to non-powerlaw scaling. Hence, for the first time, we demonstrate that heating rates may represent a non-universal, drive-dependent property of the dynamics. This appears at odds with the common paradigm that heating rates can be estimated in a model-independent way~\cite{abanin_15,mori_15}, and presents a new challenge for the theoretical understanding of (pre-) thermalization in Floquet systems. 

The asymptotic character of the inverse-frequency expansion has been shown to cause its failure to capture the onset of heating in the unconstrained thermalization stage of the dynamics~\cite{bukov_15_res}. In contrast to common beliefs, we show that it is the expansion Hamiltonian $H_\mathrm{eff}$, that the system thermalizes to, past the prethermal plateau. We demonstrate that, at any fixed stroboscopic time, the system is in an (approximate) thermal state, with temperature set by the instantaneous energy density of $H_\mathrm{eff}$. The infinite temperature state is then approached with a gradually increasing temperature. 
This slow-heating dynamics offers an exciting new possibility to control temperature in \emph{isolated} quantum simulators, such as ultracold atoms, trapped ions and superconducting circuits, using a periodic drive.

\begin{figure}[t!]
	\includegraphics[width=1.0\columnwidth]{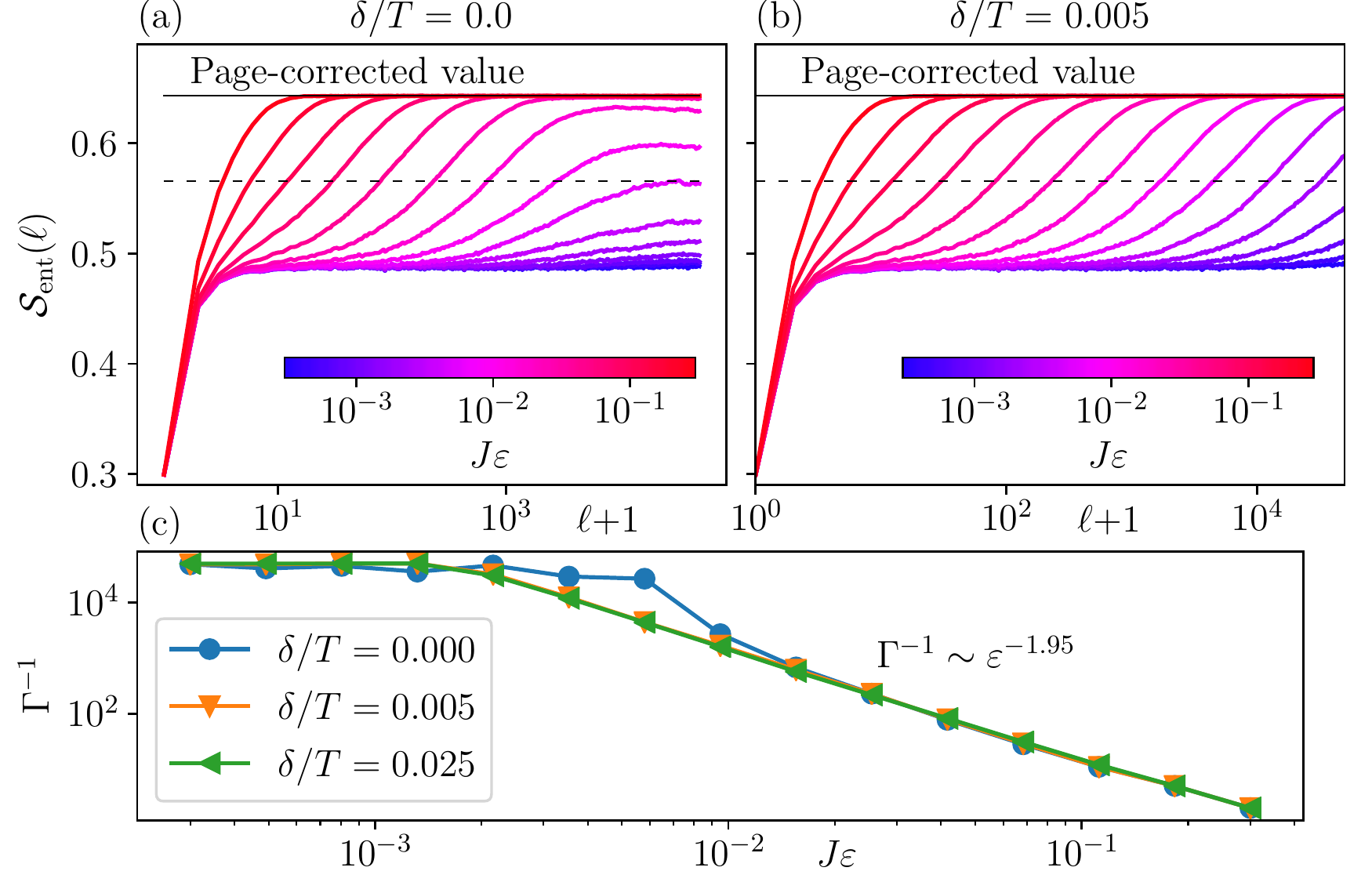}
	\caption{\label{fig:E_vs_ell_pure}
		Stroboscopic evolution of the entanglement entropy $\mathcal{S}_\mathrm{ent}$ in the $\varepsilon$-vicinity of the commensurate point $T_k^\ast$ 
		generated by the Hamiltonian $H_1$. 
		{\bf (a)} strictly periodic drive; 
		{\bf (b)} noise-perturbed drive with $\delta/T=0.005$ [see text]. 
		We display a logarithmically decreasing number of data points at large $\ell$.
		{\bf (c)} heating rate $\Gamma^{-1}$ vs.~$\varepsilon$ in the noise-perturbed and periodic drives, defined by the dashed horizontal black line. 
		We choose $15$ logarithmically spaced $\varepsilon$ values, $\varepsilon \in [3\times 10^{-4},3\times 10^{-1}]$ (the interval limits including). 
		The parameters are $h_z/J=0.809$, $h_x/J=0.9045$, $\gamma/J=1$, and $k=2$. 
		The system and subsystem sizes are $L=20$ and $L_A=10$, respectively.
}
\end{figure}

\emph{Finite Frequency Prethermalization.---}Consider the family of $T$-periodic Floquet unitaries
\begin{eqnarray}
\label{eq:UF_ast}
U_F(T) = \mathrm e^{-i T H/4}\mathrm e^{-i T V/2}\mathrm e^{-i T H/4} = \mathrm e^{-i T H_F},
\end{eqnarray}
parametrized by the drive Hamiltonian $H$ and the kick operator $V$, and subject to two constraints: 
(i) $H$ can be either integrable or nonintegrable, so long as it is local and the average Hamiltonian $H_\mathrm{ave}\!=\! H\!+\!V$ is nonintegrable;
(ii) the kick operator $V$ has a commensurate spectrum, i.e.~there exist periods $T^\ast_k$ with $k\in\mathbb{N}$ where $\exp(-i T^\ast_k V) \!=\! \mathbb{1}$. For instance, $V$ can be any short-range density-density interaction in strongly-correlated lattice models; alternatively, for this paper we focus on spin-$1/2$ systems and, without loss of generality, choose a global magnetic field.
We consider $T$-periodic stroboscopic dynamics $t=\ell T$ ($\ell\in\mathbb{N}$) with $U(\ell T)=[U_F(T)]^\ell$, generated by the exact Floquet Hamiltonian  $H_F$; in the high-frequency limit, it can be approximated by an effective Hamiltonian using an inverse-frequency expansion, $ H_F \!\approx\! H_\mathrm{eff} \!=\! H_\mathrm{ave} \!+\! \mathcal{O}(T)$.

By construction, at $T\!=\!2T^\ast_k \!=\!2 \pi k/\gamma$ with $k\in\mathbb{N}$, the dynamics of Eq.~\eqref{eq:UF_ast} defines a quench problem to the static Hamiltonian $H$: $U_F(2T^\ast_k)\!=\!\exp(-iT^\ast_k H)$, and heating is prohibited by energy conservation. 
Note that $k=0$ corresponds to the familiar infinite-frequency limit, where the Floquet dynamics features a prethermal plateau. Hence, this setup presents a natural way to generalize the concept of prethermalization to intermediate and low frequencies ($k\!>\!0$), on the order of the single-particle energy scales in $H$ and $V$. In this paper, we investigate the imminent question about the heating behavior of the family of drives from Eq.~\eqref{eq:UF_ast} in the $\varepsilon$-vicinity of the commensurate periods $T\!=\!2T^\ast_k$ with $k \! > \! 0$ [Fig.~\ref{fig:schematic}]. 
Coincidentally, our analysis is directly applicable to periodically-kicked systems~\cite{prosen_98a}:
for $T\!=\!2(T_k^\ast\!+\!\varepsilon)$, the dynamics of Eq.~\eqref{eq:UF_ast} reduces to that of the system $H$ subject to $T$-periodic kicks $V$ of strength $\varepsilon$: $U_F\left(T\!=\!2(T^\ast_k + \varepsilon)\right) \!=\!  \mathrm e^{-i T H/4}\mathrm e^{-i \varepsilon V}\mathrm e^{-i T H/4}$.

\emph{Nonintegrable Drive.---}First, let us investigate the dynamics generated by the nonintegrable spin-$1/2$ mixed-field Ising model with periodic boundary conditions
\begin{eqnarray}
\label{eq:Ising_mixed}
H_1= \sum_{j=1}^L J \sigma^z_{j+1}\sigma^z_j + h_z \sigma^z_j + h_x \sigma^x_j,\quad V = \gamma \sum_{j=1}^L \sigma^x_j.
\end{eqnarray}
The Pauli matrices obey $[\sigma^\alpha_i,\sigma^\beta_j]\!=\!2i\delta_{ij}\epsilon^{\alpha\beta\gamma}\sigma^\gamma_j$.  
We work in the zero momentum sector of positive parity, where the only local integral of motion of $H_1$ is energy. For concreteness, we prepare the system in the domain wall state $|\psi_i\rangle=\mathcal{P}|\uparrow\dots\uparrow\downarrow\dots\downarrow\rangle$ projected onto the same symmetry sector~\footnote{We checked that, our conclusions are independent of the choice of initial pure state.}.
For $T\!=\!2(T_k^\ast\!+\!\varepsilon)$ with $k\!>\!0$, $H_1$, and with it $H_\mathrm{eff}\!=\! H_1/2\!+\!\mathcal{O}(\varepsilon)$, exhibits Wigner-Dyson level-spacing statistics. Thus, since the dynamics generated by Eqs.~\eqref{eq:UF_ast} and~\eqref{eq:Ising_mixed} violates energy conservation, according to the Eigenstate Thermalization Hypothesis (ETH), we expect to observe thermalizing dynamics~\cite{dalessio2016quantum}.

We compute the exact evolution of the system, $|\psi(\ell)\rangle= U_F^\ell |\psi_i\rangle$, numerically at stroboscopic times $\ell T$ up to $5\times10^4$ driving cycles. Since we are interested in observable-independent features of the dynamics, we focus on the entanglement entropy density $\mathcal{S}_\mathrm{ent}(\ell) = -\frac{1}{L_A}\mathrm{tr}_A\rho^A\log\rho^A$, with the reduced density matrix (RDM) of subsystem $A$, $\rho^A=\mathrm{tr}_{\bar A}|\psi(\ell)\rangle\langle\psi(\ell)|$. Figure~\ref{fig:E_vs_ell_pure}a shows that a qualitatively similar behavior to the familiar infinite-frequency point ($k\!=\!0$), occurs in the neighborhood of the commensurate points $T^\ast_k$ with $k\!>\!0$. Indeed, for sufficiently small $\varepsilon$, we observe all four stages of thermalization [Fig.~\ref{fig:schematic}]. Infinite-temperature finite size effects are taken into account using the Page correction~\cite{Page1993}. 
In particular, prethermalization occurs in the $\varepsilon$-vicinity of the commensurate point $T^\ast_k$ away from the infinite-frequency point $k\!=\!0$. Observing the prethermal physics in $\mathcal{S}_\mathrm{ent}$, we anticipate that this behavior is generic.

The duration of the prethermal plateau is parametrically controlled by the deviation  $\varepsilon$ from the commensurate point [Fig.~\ref{fig:schematic}]. To quantify the dependence, we define the heating time $\Gamma^{-1}(\varepsilon)$ as the duration at which the entropy curves reach half the value between the prethermal plateau and the Page-corrected maximum entropy. The heating rate $\Gamma(\varepsilon)$  corresponds to the inverse heating time. 
Recall that, for $k\!=\!0$, heating times are exponentially long~\cite{abanin_15,mori_15}, i.e.~$\Gamma^{-1}\!\propto\!\exp(\xi/\epsilon)$\footnote{Up to a logarithmic correction in one dimension~\cite{Avdoshkin2020}.}. 
In contrast, for the dynamics generated by $H_1$ at $k\! >\! 0$, here we find algebraically suppressed heating $\Gamma^{-1} \!\propto\! \varepsilon^{-\alpha}$ with $\alpha\!\approx\! 2$ [Fig.~\ref{fig:E_vs_ell_pure}c], in accord with Fermi's Golden Rule~\cite{mallaya2019heating}. Despite the established belief that this behavior is generic, we will see that heating rates are, in fact, model-dependent. Curiously, for $k\!>\!0$, we find no change in the heating times with $k$~\cite{fleckenstein_long}.  

The Golden Rule heating exponent indicates a fully ergodic dynamics. Yet, a careful examination of the simulation data suggests  that, despite the nonintegrability of $H_\mathrm{eff}$, the dynamics is not completely ergodic out to the very long times. 
Indeed, the entanglement curves for some values of $\varepsilon$ do not reach infinite temperature after leaving the prethermal plateau. This implies that the system does not explore the entire available Hilbert space ergodically [Fig.~\ref{fig:E_vs_ell_pure}a]; instead it gets stuck in a non-thermal steady state. We verified that this peculiar feature is a finite size effect \cite{fleckenstein_long}. Yet, we report a novel procedure which allows us to conveniently remove it
at finite system sizes by perturbing the periodicity of the drive, e.g.~by adding to the duration of the Hamiltonian $H$ a small number $\delta$ sampled uniformly at random from $[0,0.05T]$ at every drive cycle: 
$U_F\to U_\delta\left(T\!=\!2(T^\ast_k + \varepsilon)\right) =  \mathrm e^{-i (T+\delta) H/4}\mathrm e^{-i \varepsilon V}\mathrm e^{-i (T+\delta) H/4}$. Notably, despite breaking the periodicity, noise-perturbed drives do not remove or shrink the prethermal plateau for sufficiently small $\delta$ [Fig.~\ref{fig:E_vs_ell_pure}b,c]. This is expected for truly ergodic/Markovian dynamics; it can be understood formally by noticing that the leading order effective Hamiltonian, $H_1/2$, only acquires a multiplicative correction of $[1+\delta/(2T)]$~\cite{fleckenstein_long}.

\begin{figure}[t!]
	\includegraphics[width=0.48\textwidth]{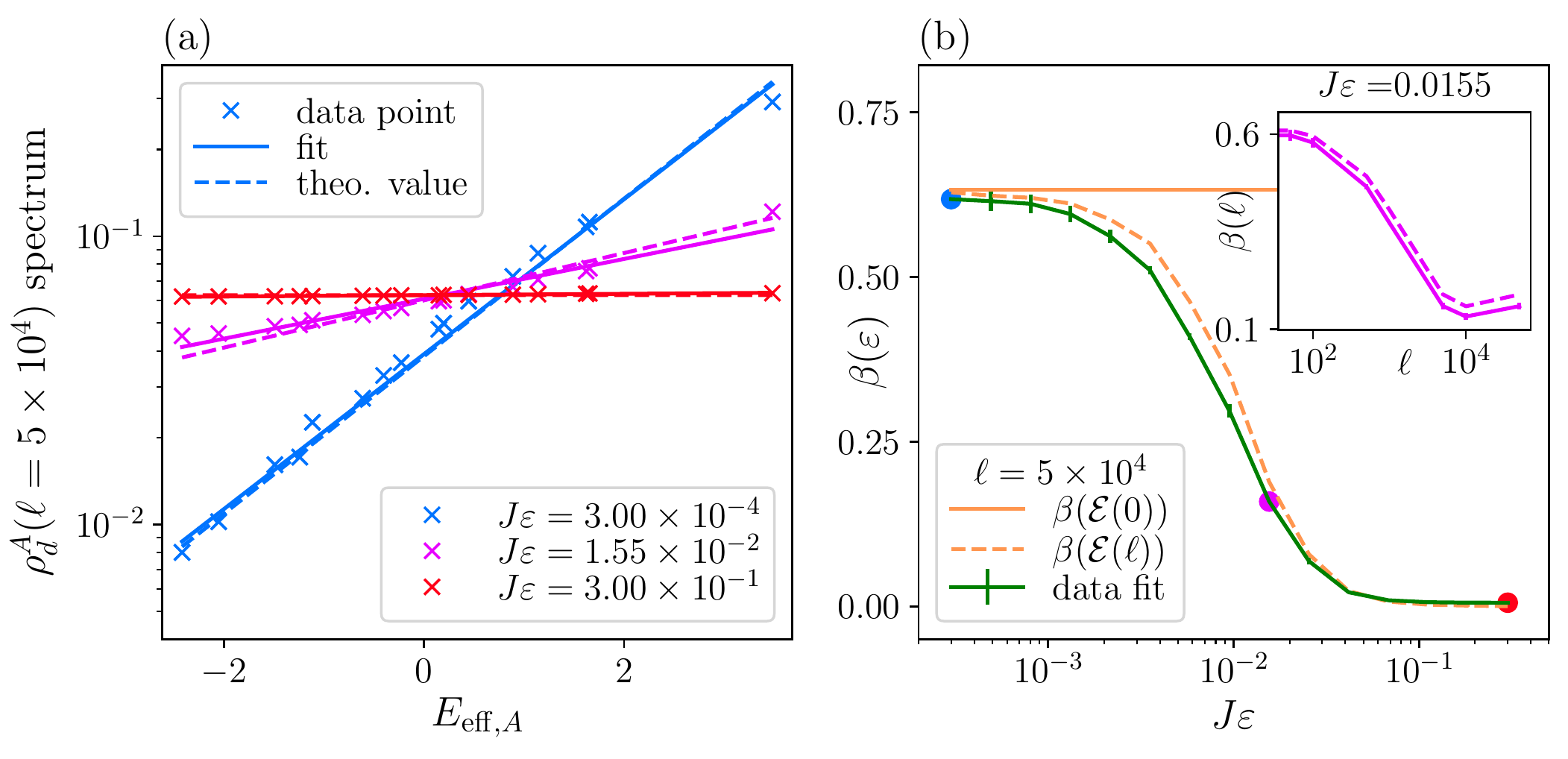}
	\caption{\label{fig:beta_pure} 
		{\bf (a)} Spectrum of the RDM (crosses) against the eigenvalues of $H^A_\mathrm{eff}\approx H^A_1/2$ for three different values of $\varepsilon$; the dashed line indicates the solution for $\beta$ from Eq.~\eqref{eq:beta_of_E}. 
		{\bf (b)} $\beta$ as a function of $\varepsilon$. The solid green line with error bars marks the values extracted from the fits in (a) at $\ell=5\times10^4$. The error bars constitute the uncertainty of the least square fit; 
		the solid orange line is the ETH prediction for the prethermal plateau; the dashed orange line is the solution to Eq.~\eqref{eq:beta_of_E} using the instantaneous energy density $\mathcal{E}(\ell)$ at $\ell=5\times10^4$. Filled circles indicate the three values of $\varepsilon$ shown in (a). The inset shows the same quantities as a function of $\ell$.	
		We used $L_A=4$ and $M=20$; the model parameters are the same as in Fig.~\ref{fig:E_vs_ell_pure}.
	}
\end{figure}

\begin{figure*}[t!]
	\includegraphics[width=1\textwidth]{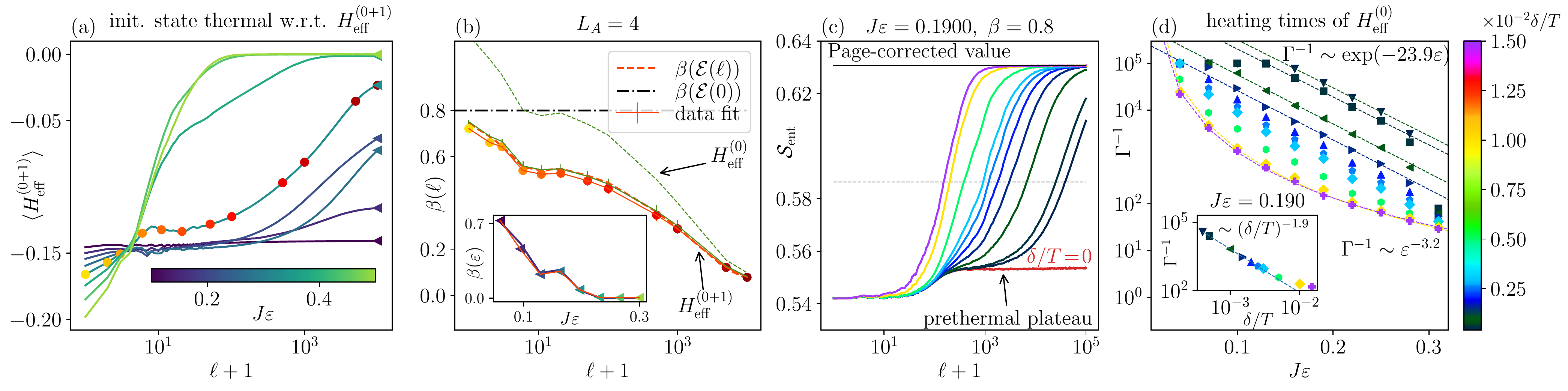}
	\caption{\label{fig:thermal_case_2a} 
		{\bf (a)} Stroboscopic time evolution for different $J\varepsilon$ ($J=0.6$) with the initial state thermal w.r.t. $H_{\mathrm{eff}}^{(0+1)}$ at $\beta=0.8$. {\bf (b)} $\beta$ as a function of $\ell$: the solid red line with error bars marks the values extracted from the spectrum of the RDM for a subsystem $L_A=4$ from the evolution in (a) using the eigenvalues of $H_{\mathrm{eff}}^{(0+1)}$($E_A^{(0+1)}$) (colored dots serve as a guide to the eye). The red dashed line is the solution to Eq.~\eqref{eq:beta_of_E}, i.e. $\beta(\mathcal{E}(\ell))$ (using $E_A^{(0+1)}$), with periodic boundary conditions on the subsystem~\cite{fleckenstein_long}. The green lines show ETH-predicted (dashed) and fitted (solid with errorbars) values of $\beta(\ell)$ derived by using the eigenvalues of $H_{\mathrm{eff}}^{(0)}$; the inset shows $\beta(\varepsilon)$ at $\ell=10^4$.
		{\bf (c)} Stroboscopic time evolution of the entanglement entropy (initial state thermal w.r.t $H_{\mathrm{eff}}^{(0)}$ at $\beta=0.8$ with $J=1$) at $J\varepsilon=0.19$ for different perturbation values $\delta/T$. 
		{\bf (d)} Heating times extracted from the data in (c) for different $\delta/T$. The inset shows the scaling of heating times as a function of $\delta/T$. 
		The remaining parameters are as in Fig.~\ref{fig:E_vs_ell_pure}, except for $L=16$ in (c) and (d).
	}
\end{figure*}

Although the prethermal plateau is clearly discernible in the dynamics, its existence does not immediately imply the thermal property of the underlying state. 
Since the effective Hamiltonian is nonintegrable already to leading order in $\varepsilon$, motivated by ETH we expect the RDM $\rho^A$ to evolve into a thermal state $\rho_\mathrm{th}^A$~\cite{garrison2018does, dymarsky2018subsystem} with temperature, corresponding to the energy density of the initial state~\cite{dalessio2016quantum,deutsch2018eigenstate}. To demonstrate the applicability of ETH we solve the implicit equation
\begin{equation}
\label{eq:beta_of_E}
\mathcal{E}_i = \frac{1}{L_A}\mathrm{tr}_A \left(\rho_\mathrm{th}^A  H_\mathrm{eff}^A\right),\quad 
\rho_\mathrm{th}^A = \mathrm{e}^{-\beta H_\mathrm{eff}^A} \big/ \mathrm{tr}_A \mathrm{e}^{-\beta H_\mathrm{eff}^A}
\end{equation}
for $\beta$, where $\mathcal{E}_i \!=\! \langle\psi_i|H_\mathrm{eff}|\psi_i\rangle/L$ is the initial energy density. This provides us with a theoretically predicted reference value for the inverse (prethermal) temperature $\beta$. We can now compare it against the value we obtain from the numerical data: we can extract a temperature from the spectrum of the reduced diagonal density matrix $\rho_d^A(\ell)\!=\!\mathrm{tr}_{\bar A}\rho_d(\ell)$, where $\rho_d \!\approx\! \frac{1}{M}\sum_{m=0}^{M} |\psi(\ell\!+\!m)\rangle\langle\psi(\ell\!+\!m)|$ is the density matrix of the diagonal ensemble~\cite{polkovnikov2011colloquium}.

Fig.~\ref{fig:beta_pure}a indicates that, starting from a pure state of the full system, the subsystem evolves into a (approximate) thermal state, whose temperature matches well the value predicted by ETH w.r.t.~$H_\mathrm{eff}$~\cite{fleckenstein_long}. In particular, for $\varepsilon\lesssim10^{-3}$, the long-lived prethermal plateau appears to be well described by a thermal density matrix with inverse temperature $\beta(\mathcal{E}_i)$. 
Remarkably, for the first time, the data allow us to make predictions beyond those of ETH for the prethermal plateau: When the system starts heating up and leaves the prethermal plateau, its state at subsequent times is still well-described (to an excellent precision) by a thermal state w.r.t.~the approximate $H_\mathrm{eff}$ whose temperature is set by the instantaneous energy density  $\mathcal{E}(\ell)$ [cf.~dashed and solid lines, Fig.~\ref{fig:beta_pure}b]. This comes as a surprise, since heating processes emerge due to non-analytic terms present in $H_F$ but missing from $H_\mathrm{eff}$ to any order in the inverse-frequency expansion~\cite{bukov_15_res}. Consequently, $H_\mathrm{eff}$ is not capable of predicting the value of $\mathcal{E}(\ell)$ past the prethermal regime. Nonetheless, given $\mathcal{E}(\ell)$ and $H_\mathrm{eff}$ one can reconstruct the thermal state of the system at the stroboscopic time $\ell$. 
By simulating the dynamics of thermal initial states at different temperatures, we ruled out any energy-density dependence of this effect~\cite{fleckenstein_long}.

\emph{Integrable Drive.---}While nonintegrable drives immediately lift energy conservation and unlock thermalizing dynamics, for integrable drives $H$ the same small parameter $\varepsilon$ breaks both energy conservation and the integrability of $H_\mathrm{eff}$. An example of such a system is the paradigmatic Floquet-Ising chain that recently emerged as a convenient model to study nonequilibrium phenomena~\cite{Talia2019},
\begin{eqnarray}
\label{Eq:Ising3}
H_3= \sum_{j=1}^L J \sigma^z_{j+1}\sigma^z_j+ h_z \sigma^z_j,\quad V = \gamma \sum_{j=1}^L \sigma^x_j.
\end{eqnarray}
Although the zeroth-order approximation to $H_F(\varepsilon)$ is integrable, higher-order terms contain nonintegrable corrections. Applying the replica inverse-frequency expansion~\cite{vajna2018replica} to resum a Baker-Campbell-Hausdorff subseries~\footnote{
	The resummation based on the replica expansion requires a two-step drive as opposed to the symmetric three-step drive in Eq.~\eqref{eq:UF_ast}. For the remainder of the paper we thus discuss the two-step drive: $U_F(T) = \mathrm e^{-i T H/2}\mathrm e^{-i T V/2}$~\cite{fleckenstein_long}; we verified that our conclusions do not depend on this gauge choice.}, 
allows us to analytically obtain a closed-form expression for the first-order correction: $H_{\mathrm{eff}}^{(0+1)}(\varepsilon)\!=\!H_{\mathrm{eff}}^{(0)}\!+\!H_{\mathrm{eff}}^{(1)}(\varepsilon)$ [cf.~Ref.~\cite{fleckenstein_long} for the expression]. 
In contrast to the dynamics generated by $H_1$, the first-order correction is indeed decisive to properly capture thermalization w.r.t.~$H_3$, as it is observed for relatively large values of $\varepsilon \!\sim\!  10^{-1}$ [Fig.~\ref{fig:thermal_case_2a}].
To substantiate this claim, we initialize the system in a thermal state w.r.t.~$H_{\mathrm{eff}}^{(0+1)}$, and evolve it according to the exact protocol of Eq.~\eqref{eq:UF_ast}~\footnote{In fact, we apply extra initial and final rotations as the replica expansion is formulated for two-step drives~\cite{fleckenstein_long}.}. Thermal states allow us to avoid any initial state dependence that might spoil thermalization as $\varepsilon\rightarrow 0$. Since simulating exact thermal states is infeasible for the system sizes of interest, we resort to quantum typicality to approximate a thermal state by an ensemble of pure states~\cite{bartsch2009dynamical,reimann2018dynamical,reimann2019typicality,richter2019combining, fleckenstein_long}. 
We then use the instantaneous energy density $\mathcal{E}(\ell)$ to solve Eq.~\eqref{eq:beta_of_E} and compare the obtained $\beta$ against the value fitted from the spectrum of the reduced diagonal density matrix. 
Fig.~\ref{fig:thermal_case_2a}(a-b) shows that the prethermal physics in the vicinity of the commensurate point $T^\ast_k$ admits an approximate analytical description. Moreover, we verified that including the first correction, $H_{\mathrm{eff}}^{(0+1)}$, results in a more accurate description of the thermalizing dynamics [cf.~Fig.~\ref{fig:thermal_case_2a}b, dashed and solid green lines]. 
Hence, the ability of the inverse-frequency expansion (supplemented with the instantaneous energy density) to capture the physics of the system past the prethermal plateau, applies equally to integrable and nonintegrable drives. Note that this cannot be interpreted as thermalization w.r.t.~$H_3$ alone which is applied for $T\gg\varepsilon$, since $H_3$ and $V$ are both integrable.  

Surprisingly, we find that the heating rates of the dynamics generated by $H_3$ do not obey a power-law scaling with $\varepsilon$. Applying an infinitesimal perturbation $\delta$ to maintain ergodicity in the dynamics at long times, the duration of the prethermal plateau increases exponentially with $\varepsilon$ over at least two decades [Fig.~\ref{fig:thermal_case_2a}d], in contrast to the Fermi's Golden Rule behavior observed for $H_1$. We emphasize that this exponential, $\Gamma^{-1}\!\propto\!\exp(-\xi\varepsilon)$, appears different from the scaling behavior $\Gamma^{-1}\!\propto\!\exp(\xi/\varepsilon)$ close to the infinite-frequency point $k\!=\!0$, and thus, we do not expect it to survive as $\varepsilon\!\to\! 0$. To exhibit the difference of the perturbation-free periodic evolution at $\delta\!=\!0$ with the Golden Rule prediction, we consider stronger perturbations, and show that the heating rate dependence crosses over to a powerlaw, albeit with an exponent larger than $2$. Thus, unlike for the nonintegrable drive $H_1$, here we find a clear dependence of the prethermal plateau duration on $\delta$. 
Interestingly, the absence of a Golden Rule scaling implies a non-Markovian dynamics and thus the state retains some memory of its evolution; in turn, this suggests the possibility for a synchronization effect caused by the periodic dynamics of $H_3$, which opens up new avenues for further investigation, e.g.~considering disordered kick strengths $\gamma\to\gamma_j$, etc. At the same time, our results show a clear heating rate dependence on the drive model. This raises a question about the existence of improved heating rate estimates, tailored for specific families of Floquet drives. 

\emph{Outlook.---}Extending prethermal Floquet physics beyond the high-frequency regime lays the foundations for novel generalizations of Floquet engineering in the low-frequency regime~\cite{rodriguez2018floquet}. They can be useful in experimental platforms where the existence of higher bands or many-body processes renders the high-frequency limit inaccessible. A concrete example would be the observation of prethermal time crystalline behavior around the commensurate points $T^\ast_k$ for $k\!>\!0$~\cite{pizzi2020time} .  
Moreover, experiments with ultracold atoms~\cite{singh2019quantifying,abadal2020floquet} can shed light on the behavior of large system sizes. The slow thermal dynamics, captured by the inverse-frequency expansion beyond the prethermal plateau, offers an exciting new possibility to tune temperature in isolated quantum simulators, without access to a conventional thermal bath. This can be used, e.g., to trigger and observe thermal phase transitions or study the interplay between thermal and quantum criticality in a controlled way using Floquet drives.

\emph{Acknowledgments.---}We wish to thank J.~Bardarson, A.~Das, W.~W.~Ho,  F.~Huveneers, V.~Khemani, A.~Polkovnikov, F.~Pollmann, T.~Prosen, W.~De Roeck, B.~Trauzettel and P.~Weinberg for valuable discussions. 
C. F. acknowledges financial support from the DFG (SPP1666, SFB1170 ToCoTronics), the Wüzburg-Dresden Cluster of Excellence ct.qmat, EXC2147, project-id 39085490, the Elitenetzwerk Bayern Graduate School on Topological insulators and the ERC Starting Grant No. 679722.
M.B.~was supported by the U.S. Department of Energy, Office of Science, Office of Advanced Scientific Computing Research, under the Accelerated Research in Quantum Computing (ARQC) program, the U.S. Department of Energy under cooperative research agreement DE-SC0009919, the Emergent Phenomena in Quantum Systems initiative of the Gordon and Betty Moore Foundation, and the Bulgarian National Science Fund within National Science Program VIHREN, contract number KP-06-DV-5
This research was supported in part by the International Centre for Theoretical Sciences (ICTS) during a visit for participating in the program -  Thermalization, Many body localization and Hydrodynamics (Code: ICTS/hydrodynamics2019/11).
We used \href{https://github.com/weinbe58/QuSpin#quspin}{Quspin} for simulating the dynamics of the quantum systems~\cite{weinberg2017quspin,weinberg2019quspin}.
The authors are pleased to acknowledge that the computational work reported on in this paper was performed on the Shared Computing Cluster which is administered by Boston University’s Research Computing Services and on the W\"urzburg HPC cluster.

\bibliographystyle{apsrev4-1}
\bibliography{./bibliography}

\end{document}